\documentstyle[12pt]{article}
\def\NPB#1#2#3{Nucl. Phys. B {\bf#1} (19#2) #3}
\def\PLB#1#2#3{Phys. Lett. B {\bf#1} (19#2) #3}

\def\PRD#1#2#3{Phys. Rev. D {\bf#1} (19#2) #3}
\def\PRL#1#2#3{Phys. Rev. Lett. {\bf#1} (19#2) #3}

\def\ARNP#1#2#3{Ann. Rev. Nucl. Part. Sci. {\bf#1} (19#2) #3}

\begin{document}
\newcommand{\newc}{\newcommand}
\newc{\sm}{Standard Model}
\newc{\dac}{discrete anomaly cancellation }
\date{}
\title{Fermion masses and mixing angles from gauge symmetries.}
\author{Luis Ib\'a\~nez,\\ Departamento de Fisica Te\'orica \\
                   Universidad Aut\'onoma de Madrid \\
Cantoblanco, 28034 Madrid\\
\\
\centerline{and} \\
   \\Graham G.
Ross\thanks{SERC Senior
Fellow},\\Department of Physics,\\
Theoretical Physics,\\
University of Oxford,\\
1 Keble Road,\\
Oxford OX1 3NP}
\maketitle
\vspace{-6in}\hspace{4in}OUTP-9403; FTUAM94/7 \vspace{6in}
\begin{abstract}
The structure of the quark and lepton masses and mixing angles
provides one of the few windows we have on the underlying physics
generating the \sm. In an attempt to identify the underlying
symmetry group we look for the simplest gauge extension
of the SUSY standard model capable of generating the observed
structure. We show that the texture structure and hierarchical
form found in the (symmetric) quark and lepton mass matrices
follows if one extends the gauge group of the standard model to
include an  horizontal $U(1)$ gauge factor, constrained by the
need for anomaly cancellation. This $U(1)$ symmetry is
spontaneously broken slightly below the unification/string scale
leaving as its only remnant the observed structure of masses and
mixings. Anomaly cancellation is possible only in the context of
superstring theories via the Green Schwarz mechanism with
$sin^2(\theta_W)=3/8$.

\end{abstract}
\maketitle

\newpage

\section{Introduction}
There has been considerable success explaining the parameters of
the \sm \ in the framework of a supersymmetric extension of the
model with a stage of unification. The measured values of the
gauge couplings are consistent with their normal
unification     values with a unification scale of
O($10^{16}GeV$) provided, at      low energies $\le O(10^3GeV)$,
the \sm \ spectrum is extended to      that of the  minimal
supersymmetric model (the MSSM)\cite{9e}. In addition
     the pattern (and magnitude) of spontaneous breakdown of the
\sm \    follows naturally from the structure of radiative
corrections in                 the MSSM provided there is some
unification of the supersymmetry breaking masses at the
unification scale\cite{9e,9f}.
This simplicity in the parameters of the (supersymmetric) \sm \
a   high scales appears to extend to some of the couplings
involved  in determining the fermion masses. The measured values
of the     bottom quark and the $\tau$ lepton are consistent with
their       equality at the unification scale\cite{1,2}. Further
the mixing angles and     masses have values consistent with the
appearance of ``texture''     zeros in the mass
matrix\cite{3,4,5,5e}, such texture zeros indicating the
appearance of additional symmetries beyond the \sm.

In this paper we will explore the possibility that at least some
of the symmetries giving rise to this texture structure are new
gauge symmetries. Given the success of the MSSM we will look for
the  minimal extension of the MSSM able to generate a Yukawa
structure which is phenomenologically viable. We consider only
the case of symmetric mass matrices\footnote{This restriction is
also motivated by our desire to identify the maximally symmetric
possibility consistent with observation, and suggests there is
a further left-right symmetry at high scales.} for this allows
us to determine the structure of the mass matrices with texture
zeros and hence provides us with a definite starting point for
the search for new symmetries. Surprisingly we find that the
simplest possibility in which one extends  the
\sm \ to include an extra $U(1)$ symmetry is already sufficient
to generate an interesting texture structure.
However, this extra $U(1)$ has  a stringy nature: by itself it
is anomalous and its anomaly is cancelled by a Green-
Schwarz\cite{6} mechanism coming from an underlying string
theory. Remarkably this cancellation only works if the Kac Moody
level structure associated with the \sm \ gauge groups is such
that the gauge couplings have their normal unification values
leading to the successful unification predictions noted above.

\section{Quark mass matrices.}
\label{sec:qmm}

The structure of the quark mass matrices is not directly measured
because the charged weak current only tells us about the mixing
in the left-handed sector. If, however,  one adds the requirement
of symmetry relating the left to the right sector, the extraction
of the mass matrices from data becomes feasible\cite{7}. In
particular we wish to consider the possibility that the mass
matrices are left-right symmetric
i.e., invariant under the $Z_2$ exchange
$q_R\leftrightarrow q_L$. In this case the experimental
measurements of quark masses and mixing angles may be used to
determine the mass matrices with the maximum number of texture
zeros. As we shall discuss, given the form of these matrices
allows us to determine whether additional symmetries are capable
of generating their structure and hence of giving the observed
quark masses and mixing angles.

\begin{table}
\centering
\begin{tabular}{|c|c|c|} \hline
Solution & ${\bf Y}_u$ & ${\bf Y}_d$ \\ \hline
1 & $\left(
\begin{array}{ccc}

0 & \sqrt{2}\lambda^6 & 0 \\
\sqrt{2}\lambda^6 & \lambda^4 & 0 \\
0 & 0 & 1
\end{array}
\right)$ & $\left(
\begin{array}{ccc}

0 & 2\lambda^4 & 0 \\
2\lambda^4 & 2\lambda^3 & 4\lambda^3 \\
0 & 4\lambda^3 & 1
\end{array}
\right)$
\\ \hline

2 & $\left(
\begin{array}{ccc}

0 & \lambda^6 & 0 \\
\lambda^6 & 0 & \lambda^2 \\
0 & \lambda^2 & 1
\end{array}
\right)$ & $\left(
\begin{array}{ccc}

0 & 2\lambda^4 & 0 \\
2\lambda^4 & 2\lambda^3 & 2\lambda^3 \\
0 & 2\lambda^3 & 1
\end{array}
\right)$
\\ \hline

3 & $\left(
\begin{array}{ccc}

0 & 0 & \sqrt{2}\lambda^4 \\
0 & \lambda^4 & 0 \\
\sqrt{2}\lambda^4 & 0 & 1
\end{array}
\right)$ & $\left(
\begin{array}{ccc}

0 & 2\lambda^4 & 0 \\
2\lambda^4 & 2\lambda^3 & 4\lambda^3 \\
0 & 4\lambda^3 & 1
\end{array}
\right)$
\\ \hline

4 & $\left(
\begin{array}{ccc}

0 & \sqrt{2}\lambda^6 & 0 \\
\sqrt{2}\lambda^6 & \sqrt{3}\lambda^4 & \lambda^2 \\
0 & \lambda^2 & 1
\end{array}
\right)$ & $\left(
\begin{array}{ccc}

0 & 2\lambda^4 & 0 \\
2\lambda^4 & 2\lambda^3 & 0 \\
0 & 0 & 1
\end{array}
\right)$
\\ \hline

5 & $\left(
\begin{array}{ccc}

0 & 0 & \lambda^4 \\
0 & \sqrt{2}\lambda^4 & \frac{\lambda^2}{\sqrt{2}} \\
\lambda^4 & \frac{\lambda^2}{\sqrt{2}} & 1
\end{array}
\right)$ & $\left(
\begin{array}{ccc}

0 & 2\lambda^4 & 0 \\
2\lambda^4 & 2\lambda^3 & 0 \\
0 & 0 & 1
\end{array}
\right)$
\\ \hline
\end{tabular}
\caption{Approximate forms for the symmetric textures of quark
masses. }
\label{table:1}
\end{table}

\begin{table}
\centering
\begin{tabular}{|c|c|c|c|c|} \hline
$Solution $&$ \frac{m_c}{m_t} $&$ \frac{m_u}{m_c}  $&$ V_{cb} $&$
\frac{V_{ub}}{V_{cb}}$\\ \hline
``Experiment'' & 0.006-0.01 & 0.003-0.005 & 0.02-0.05 & 0.05-0.13
\\ \hline
1 & 6.7  $10^{-3}$ & 0.0046 & 6  $10^{-2}$ & 0.068 \\ \hline 2
& 6.7  $10^{-3}$ & 0.0023 & 3.8  $10^{-2}$ & 0.0484 \\ \hline 3
& 6.7  $10^{-3}$ & 0.0046 & 6  $10^{-2}$ & 0.078 \\ \hline 4 &
4.9  $10^{-3}$ & 0.0087 & 6.8  $10^{-2}$ & 0.040 \\ \hline 5 &
6.1  $10^{-3}$ & 0.003 & 4.8  $10^{-2}$  & 0.068 \\ \hline
\end{tabular}
\caption{Predictions following from the five symmetric texture
solutions using $\lambda$=0.22.  All solutions give $V_{us}$=0.22
and $\frac{m_d}{m_s}$=0.05 and $\frac{m_s}{m_b}$=0.03, in
agreement with the experimental results $\frac{m_d}{m_s}$=0.04-
0.067 and $\frac{m_s}{m_b}$=0.03-0.07[10].}
\label{table:4}
\end{table}

The symmetric mass matrices with five texture
zeros\footnote{Although six texture zeros is the maximum
possible, no examples were found consistent with the masses and
mixing angles.} are given in Table \ref{table:1}\cite{7}. In
Table \ref{table:1} $\lambda$ is an expansion parameter,
essentially the (1,2) matrix element of the CKM matrix.  We
should stress that these are approximate forms in the sense that
the texture zeros should be interpreted as being zero only up to
the order that does not significantly change the masses and
mixing angles. For example Solutions 1, 2 and 4 all have texture
zeros in the (1,1) and (1,3) positions of both $Y_u$ and $Y_d$.
This leads to the predictions\cite{7}
\begin{eqnarray}
\mid \frac{V_{ub}}{V_{cb}}\mid  & = &  \sqrt{\frac{m_u}{m_c}}
\nonumber \\
\mid V_{us} \mid & = & (\frac{m_d}{m_s}+ \frac{m_u}{m_c}
+2\sqrt{\frac{m_d m_u}{m_s m_c}} \cos{\phi'})^{\frac{1}{2}}
\nonumber \\
\label{eq:tz}
\end{eqnarray}
where  $\phi'$ is a phase related to the phases (not displayed)
of the non-zero matrix elements of Table\ref{table:1}.
To preserve these phenomenologically successful relations
requires $Y_u(1,1)<O(\lambda^8), \; Y_u(1,3)<O(\lambda^4), \;
Y_d(1,1)<O(\lambda^4), \; Y_d(1,3)<O(\lambda^2)$. On the other
hand the (2,2) zero in $Y_u$ in Solution 2 is not so critical and
an entry of $O(\lambda^4)$ is acceptable.
We shall see that this is important in the models presented below
which generate {\it approximate} texture zeros.

The structure of Table \ref{table:1} strongly suggests an
underlying chiral symmetry broken by terms of $O(\lambda )$. In
the limit the symmetry is exact only the third generation is
massive and all mixing angles are zero. Symmetry breaking terms
gradually fill in the mass matrices generating an hierarchy of
mass scales and mixing angles\cite{5,8,9,10,10e}. In the next
section we address the question whether a spontaneously broken
gauge symmetry is capable of generating this structure.

\section{Gauging a family $U(1)$ symmetry}
\label{sec:gfs}

We wish to discuss the possibility that this structure results
from the simplest possible gauge extension of the \sm \, namely
an abelian horizontal gauge factor acting on the family or
generation indices. Without loss of generality we write the U(1)
in the form

\begin{equation}
U(1) =U(1)_{FD} + U(1)_{FI}
\label{eq:4}
\end{equation}
where $U(1)_{FD}$ is a family dependent symmetry, by definition
acting only on the quarks and leptons and $U(1)_{FI}$ is a family
independent symmetry. If the fermion mass matrix is to be
symmetric $U(1)_{FD}$ must act the same way on left- and right
handed components while $U(1)_{FI}$ is not constrained.  It
proves convenient to consider first the structure of $U(1)_{FD}$
as it determines the relative magnitudes of the matrix elements
within a single (up, down or lepton) mass matrix.

The $U(1)_{FD}$ charges of the MSSM states are given in Table
\ref{table:2a}. The condition of symmetric matrices requires that
all quarks(leptons) of the same i-th generation transform with
the same charge $\alpha _i(a_i)$. Through a choice of $U(1)_{FI}$
we may make $U(1)_{FD}$ traceless without any loss of generality.
Thus $\alpha_3=-(\alpha_1+\alpha_2)$ and $a_3=-(a_1+a_2)$.
\begin{table}
\begin{center}
\begin{tabular}{|c |ccccccc|}\hline
   & Q & u & d & L & e & $H_2$ & $ H_1$   \\
\hline
  $U(1)_{FD}$ & $\alpha _i$ & $\alpha _i$ & $\alpha _i$  & $a_i$
& $a_i$ & $-2\alpha _1$ &  $w\alpha _1$
\\
\hline
\end{tabular}
\end{center}
\caption{ $U(1)_{FD}$ symmetries. $w=-2$
corresponds to up-down symmetric matrices}
\label{table:2a}
\end{table}

With this we may now consider the constraints on the Yukawa
couplings generating the mass matrices. The $U(1)_{FD}$ charge
of the quark-antiquark pair has the form
\begin{eqnarray}
\left(
\begin{array}{ccc}
-2(\alpha_1+\alpha_2) & -\alpha_1 & -\alpha_2 \\
-\alpha_1 & 2\alpha_2 & \alpha_1 + \alpha_2 \\
-\alpha_2 & \alpha_1 + \alpha_2 & 2\alpha_1
\end{array}
\right)
\label{eq:p}
\end{eqnarray}
This matrix neatly summarises the allowed Yukawa couplings for
a Higgs boson coupling in a definite position should have charge
minus that shown for the relevant position.

For the leptons we have a similar structure of lepton-antilepton
charges
\begin{eqnarray}
\left(
\begin{array}{ccc}
-2(a_1+a_2) & -a_1 & -a_2 \\
-a_1 & 2a_2 & a_1 + a_2 \\
-a_2 & a_1 + a_2 & 2a_1
\end{array}
\right)
\label{eq:pl}
\end{eqnarray}

We will now consider in turn the implications of this structure
for the up and down quark and lepton mass matrices respectively.

\section{The up quark mass matrix.}

We first consider the up quark mass matrix. We assume that the
light Higgs, $H_2$, has $U(1)$ charge so that  only the
(3,3)    renormalisable Yukawa coupling to
$H_2$ is allowed. In this way     only the (3,3)
element of the associated mass matrix will be
non-zero as desired to reproduce the leading structure of Table
\ref{table:1}. The remaining entries
must be generated when the    $U(1)$ symmetry is
broken. Suppose  singlet fields, $\theta, \;
\bar{\theta}$, with $U(1)_{FD}$ charge -1, +1 respectively
acquire     equal vacuum expectation values (vevs)
along a ``D-flat'  direction, spontaneously
breaking this symmetry\footnote{The
spontaneous breaking of gauge symmetries at high scales
in  supersymmetric theories must proceed
along such flat directions to avoid large vacuum
energy contributions from D-terms.}.
 After
this breaking all entries in the mass matrix become
non-zero. For example, the (3,2) entry appears
at     $O(\epsilon^{\mid
\alpha_2-\alpha_1 \mid} )$ because  U(1)
charge conservation allows only a
coupling       $c^c t
H_2(\theta /M_2)^{\alpha_2-\alpha_1},
\;  \alpha_2>\alpha_1$
or   $c^ct H_2(\bar{\theta} /M)^{\alpha_1-\alpha_2},\;
\alpha_1>\alpha_2$  and
we have defined
$\epsilon=(\theta/M_2)$ where $M_2$ is the unification mass scale
which governs higher dimension operators.

Further elements may be generated depending on the values of
$\alpha_1$ and $\alpha_2$, giving the structure
\begin{eqnarray}
M_u\approx \left(
\begin{array}{ccc}
\epsilon^{\mid -4\alpha _1-2\alpha _2\mid } &
\epsilon^{\mid -3\alpha_1\mid } &
\epsilon^{\mid -  \alpha_2-2\alpha_1\mid }
\\
\epsilon^{\mid -3\alpha_1\mid } &
\epsilon^{\mid 2(\alpha_2-\alpha_1)\mid } &
\epsilon^{\mid \alpha_2-\alpha_1\mid } \\
\epsilon^{\mid -\alpha_2-2\alpha_1\mid } &
\epsilon^{\mid \alpha_2-\alpha_1\mid } & 1
\end{array}
\right)
\label{eq:mu0}
\end{eqnarray}

Note that an hierarchical structure immediately appears if
$\epsilon$ is small. Further one already draw some general
conclusions for we have

\begin{equation}
M^u_{11}\ \simeq \ {{(M^u_{13})^2}\over {M^u_{33}}}\ \ ;\ \
M^u_{22}\ \simeq \ {{(M^u_{23})^2}\over {M^u_{33}}}
\label{eq:gen}
\end{equation}
independently of what specific $U(1)_{FD}$ quantum numbers one
is assuming. Thus different elements are related in a manner
remarkably consistent with Solutions 1, 2 and 4 of Table
\ref{table:1}; a texture zero in the (1,3) position is correlated
with a texture zero in the (1,1) position.

The condition that such a zero occurs depends only on the ratio
$\alpha_2/\alpha_1$. Remarkably, for the range
$\alpha_2/\alpha_1>1$, there are texture zeros in the (1,1) and
(1,3) positions just as required in Solutions 1, 2 or 4 of Table
\ref{table:1}\footnote{These are ``zeros'' in the sense discussed
above that their values do not affect the masses or mixing angles
in leading order.}! As just noted the correlation between the
(1,3) and (1,1) zeros is obvious from the first relation of
eq(\ref{eq:gen}).

It is very easy to find a simple choice of $\alpha_2/\alpha_1$
which even generates the correct order for the non-zero elements
of one of the Solutions 1, 2 or 4 of Table \ref{table:1}. If one
requires that the (1,2) and (2,3) matrix elements be in the ratio
3:1 (as needed for Solution 2 or 4) then $\alpha_2=2\alpha_1$ and
the up quark mass matrix has the form

\begin{eqnarray}
M_u\approx \left(
\begin{array}{ccc}
\epsilon^8 & \epsilon^3 & \epsilon^4 \\
\epsilon^3 & \epsilon^2 & \epsilon \\
\epsilon^4 & \epsilon & 1
\end{array}
\right)
\label{eq:mu}
\end{eqnarray}
where we have used the freedom to set $\alpha_1=1$ through a
redefinition of the parameter $\epsilon$ and $\alpha_2$ (i.e.
$\epsilon\rightarrow (\theta/M)^{\alpha_1}, \; \alpha_2
\rightarrow \alpha_2/\alpha_1)$.

 This is the structure of solutions  2 or 4 (and indeed 1) of
Table
\ref{table:1} to the accuracy of the Table! The origin of this
structure deserves some comment. The relative magnitudes of the
(1,2) and (2,3) elements comes from our assumption about the
relative magnitudes of $\alpha_1$ and $\alpha_2$. The assumption
that these charges are quantised is quite reasonable if the
$U(1)$ is embedded in some larger non-Abelian GUT or if it comes
from a 4D superstring theory. As noted above the remaining
structure, in particular the texture zeros in the (1,1) and (1,3)
positions, are predicted by the anomaly cancellation condition
which fixed the transformation properties of the first
generation. Notice that, on the other hand, textures of type 3
or 5 would be difficult to obtain in the present approach. Thus
the present simple approach gives us a hint on what type
of generic textures to consider if we want them to be generated
by a simple abelian family structure.

\subsection{The origin of the higher dimension couplings.}
\label{sec:hdc}
It is appropriate to discuss, at this stage, how
the non-renormalisable terms needed to fill in the mass matrix
may arise.  There are three possible sources.   The first
possibility is that a term such as $c^c t H_2(\theta/M)$ occurs
at the string compactification level.
In this case we expect the mass scale M to be the string scale
$M_s$ ($\approx M_{Planck})$.  In addition there may be a
systematic suppression of such terms for all higher
dimension terms of this type are accompanied by a factor of the
form $exp(-aT)$ where a is a term-dependant constant and T is a
moduli-field which sets the overall radius of compactification.

The second possible origin of the higher dimension terms is the
mixing between light and heavy Higgs states.  Consider a string
compactification which in addition to  $H_1$ and $H_2$, leaves
additional Higgs multiplets
$H_{1,2}^{a,b...},\bar{H}_{1,2}^{a,b...}$ light.  This is indeed
what happens in many compactification schemes for, in addition
to
the three generations of quarks and leptons needed for a viable
theory, there are additional vector-like pairs of quarks, leptons
and Higgs fields. For example, in Calabi-Yau compactification,
there are $(h_{2,1}-h_{1,1})$ generations, where $h_{2,1}$ and
$h_{1,1}$ are the Hodge numbers counting (2,1) and (1,1) forms
and there are $h_{1,1}$ additional pairs of conjugate
representations.  Similarly there are usually additional light
Higgs states in conjugate representations generated on
compactification.  Of course such additional states would be an
embarrassment it they remained light at low scales but, being in
conjugate representations, they may be expected to gain mass if
the gauge symmetry is spontaneously broken after
compactification through their coupling to the scalar field
acquiring a vacuum expectation value.  Thus, in any
compactification scheme with a gauge group larger than that of
the standard model, the states $H_{1,2}^{a,b},
\bar{H}_{1,2}^{a,b...}$ may be expected to
acquire masses, $M_{1,2}$ of the order of the breaking scale of
the enlarged gauge group.
However if there is more than one such breaking scale there may
be more than one source of mass so that $M_1$ and $M_2$ may
differ.

After the various stages of spontaneous breaking at the high
scale, the Higgs state left light will be a combination of all
the original Higgs fields carrying the same $SU(3)\otimes
SU(2)\otimes U(1)$ gauge quantum numbers. In particular, due to
the $ <\theta >$ vev the light Higgs may carry a mixture of U(1)
charges

\begin{equation}
H^{light}_{1,2}  \simeq H_{1,2} + \sum_{r=1}(H_{1,2}^r
\frac{<\theta>^r}{M_{1,2}^r}
+H_{1,2}^{-r}\frac{ <\bar{\theta}>^r}{M_{1,2}^r})
\label{eq:hl}
\end{equation}
where we have denoted by $H_{1,2}^r$ an Higgs field carrying U(1)
charge  r.   Clearly $H^{light}_2$ will generate all elements of
$M^u$; for example the (3,2) matrix element will arise at
$0(<\theta>/M_2)$ where $M_2$ is the mass of the heavy $H_2^{1}$
field before the $<\theta>$ vev causes it to mix with
$H^{light}$.

The third sources of the higher dimension terms in $m^u$ is
through the mixing of quark U(1) eigenstates in a manner
analogous to that just discussed for the Higgs.
In this case
\begin{equation}
q^{light} \simeq q+\sum_{r=1} q_r \frac{<\theta>^r}{M_q^r} + q_{-
r}\frac{<\bar \theta>^r}{ M_q^r}
\end{equation}
where $M_q$ is the mass of the heavy quark state in the absence
of the $<\theta>$ vev.
This will also generate non-zero entries for $m^u$; for example
the (3,2) matrix element will arise at $0(<\theta>/M_q)$.

The question which contribution dominates is model dependent and
depends on whether the appropriate $H_1^r$ or $q_r$ is left light
after compactification, on the relative magnitudes of $M_q,M_1,
M_s$ and on whether, as may happen,  the $\theta$ couples only
to the quarks or to the Higgs.

\subsection{The down quark mass matrix.}
\label{sec:down}

We have seen that it is very easy to get an acceptable form for
the up quark mass matrix using a very simple choice for
$U(1)_{FD}$. We turn now to the question whether this is
consistent with the structure required for the down quark mass
matrix. Our requirement of a symmetric mass matrix together with
$SU(2)_L$ immediately gives the $U(1)_{FD}$ charge structure of
eq(\ref{eq:p}).  The structure of the down quark mass matrix then
depends only on the $U(1)$ charge of
$H_1^{light}$.  This will be constrained by further
anomaly cancellation   conditions, principally the
$SU(2)^2U(1)$ condition, which  we will discuss in Section
\ref{sec:final}. Here we first   consider whether any
$U(1)$ charge assignment for $H_1$ leads to an
acceptable down quark mass matrix.

We focus on the interesting case $\alpha_2=2\alpha_1$ which gave
an acceptable up quark mass matrix. Assigning a $U(1)$ charge
$w\alpha_1$ to $H_1$\footnote{If $b^cb$ has a $U(1)_{FI}$ charge,
$\delta$, we assume that $H_1$ has an additional charge -
$\delta$.} leads to the following form for the down quark mass
matrix:
\begin{equation}
M_d = \left (
\begin{array}{ccc}
\bar{\epsilon}^{\mid -6+w\mid } &
\bar{\epsilon}^{\mid -1+w\mid } &
\bar{\epsilon}^{\mid -2+w\mid } \\
\bar{\epsilon}^{\mid -1+w\mid } &
\bar{\epsilon}^{\mid 4+w\mid } &
\bar{\epsilon}^{\mid 3+w\mid } \\
\bar{\epsilon}^{\mid -2+w\mid } &
\bar{\epsilon}^{\mid 3+w\mid } &
\bar{\epsilon}^{\mid 2+w\mid }
\end{array}
\right)
\label{eq:7}
\end{equation}
where $\bar{\epsilon} = (\frac{<\theta >}{M_1})$ and $M_1$ is the
appropriate scale for these higher dimension terms (cf $M_2$ for
in the case of $M_u$). Following the discussion of Section
\ref{sec:hdc}, if the dominant source of these terms is from
string compactification then we expect $M_1=M_2=M_s$ and
$\epsilon=\bar{\epsilon}$. The same is true if these terms arise
due to quark mixing when $M_1=M_2=M_q$. However if they are due
to Higgs mixing then strong violation may occur of the $SU(2)_R$
symmetry of the quark sector which was forced on us by our
assumption of left- right- symmetry. The reason is because
vectorlike pairs  $H_{1,2},\; \bar{H}_{1,2}$ left massless after
compactification must acquire their mass via a stage of
spontaneous breaking after compactification and this need not
respect $SU(2)_R$. As a result $M_1\ne M_2$ and $\epsilon \ne
\bar{\epsilon}$.

Again we see from eq(\ref{eq:7}) that an hierarchical structure
for $m_d$ automatically results. As we will discuss this can be
of an acceptable form. For example with $w=-2$ one gets a matrix
               with the same structure as
eq(\ref{eq:mu})
\begin{equation}
M_d\approx \left(
\begin{array}{ccc}
\bar{\epsilon}^8 & \bar{\epsilon}^3 & \bar{\epsilon}^4 \\
\bar{\epsilon}^3 & \bar{\epsilon}^2 & \bar{\epsilon} \\
\bar{\epsilon}^4 & \bar{\epsilon} & 1
\end{array}
\right)
\label{eq:8}
\end{equation}
As we have discussed there is no reason why $\epsilon =
\bar{\epsilon}$; if they are not there will be difference between
the up
and down mass matrices.  For a suitable choice of
$\bar{\epsilon}$ this can reproduce quite closely the second
solution of Table \ref{table:1}. We will return to a detailed
discussion of this possibility in the next Section.

Another interesting choice is $w=-3/2$ for
which $m^d$ is given by \footnote{ This
choice requires fractional charges $\pm 1/2$
for the $\theta ,{\bar \theta }$ fields.}

\begin{equation}
M_d\approx \sqrt{\bar{\epsilon}}\left(
\begin{array}{ccc}
\bar{\epsilon}^7 & \bar{\epsilon}^2 & \bar{\epsilon}^3 \\
\bar{\epsilon}^2 & \bar{\epsilon}^2 & \bar{\epsilon} \\
\bar{\epsilon}^3 & \bar{\epsilon} & 1
\end{array}
\right)
\label{eq:9}
\end{equation}

Although not of the form of any of the matrices in
Table(\ref{table:1}), this structure may be viable if the
residual Yukawa couplings associated with each entry which are
not displayed here and are implicitly assumed here to be of
$O(1)$ are not all equal in magnitude. Note that eq(\ref{eq:9})
has the interesting implication that there is a suppression of
$m_b$ relative to $m_t$ due to the overall factor of
$\sqrt{\bar{\epsilon}}$.  This factor, together with the
difference in vevs of $H_{1,2}$ can describe why $m_b << m_t$.
In   the case of $w=-2$ the difference between $m_b$ and $m_t$
must come   entirely from the difference in $H_{1,2}$
vevs.

At this point an important remark is in order. If we try to do
without $U(1)_{FI}$ in eq(\ref{eq:4}) anomaly cancellation
requires the  $U(1)_{FI}$ charge for the $H_1$ field to be $w=+2$
so that the $SU(2)^2U(1)_{FI}$ anomalies coming from the Higgs
contribution cancel. In this case the form of the d-quark mass
matrix is a  disaster, giving
\begin{equation}
M_d = \left (
\begin{array}{ccc}
\bar{\epsilon}^{4} & \bar{\epsilon} &
    1 \\
\bar{\epsilon} & \bar{\epsilon}^{6} &
\bar{\epsilon}^{5} \\
1 & \bar{\epsilon}^{5} &
\bar{\epsilon}^{4}
\end{array}
\right)
\label{eq:14}
\end{equation}
which is clearly not viable.
Another obvious alternative to achieve cancellation of anomalies
is to assume that the Higgs fields are neutral. But in this case
one needs to set $\alpha _1=0$ and anomaly cancellation yields
$\alpha _2=-\alpha _3 $. These values in turn imply that the
three generations would be degenerate in mass.

{}From this we conclude that the $U(1)_{FI}$ component is
necessary. As we shall discuss in Section \ref{sec:final} it is
then possible to find an anomaly free solution but only in a
string theory in which the
four-dimensional version of the Green-Schwarz (GS) mechanism
cancels the anomalies.

\subsection{Phenomenology of the quark mass matrices.}

As we have seen the structures emerging from the requirement
there be an additional abelian gauge symmetry can reproduce the
hierarchical structure observed. The $U(1)$ symmetry by itself
only determines the order of the matrix elements and if the
constants of proportionality are chosen as in Solution 2 of Table
\ref{table:1} the masses and mixing angles are in agreement with
experiment as may be seen in Table \ref{table:2} (Note, as
discussed in Section \ref{sec:qmm} that the (2,2) entry of $Y_u$
in solution 2 is only zero to $O(\lambda ^2)$, allowing this
solution to be viable!).  To go further it is necessary to
determine the constants of proportionality in each matrix
element.  In this section we will explore whether any of the
structures discussed above are consistent with a larger symmetry
relating these constants.

As we discussed in Section \ref{sec:hdc}, the higher dimension
contributions may arise directly in the effective theory
descending from the string, or indirectly, through mixing of the
quarks and/or Higgs. Let us consider the case with $w=-2$.  In
this case the form of the up and down quark mass
matrices is the same due to an effective $SU(2)_L\otimes SU(2)_R$
symmetry in their couplings forced on us by our assumption of
left- right- symmetric mass matrices. In order to allow for a
difference between up and down quark mass matrices we have argued
that the mixing through the Higgs sector must dominate. In this
case the expansion parameter for the up and down quark sector
differs due to the different masses of the $H_1$ and $H_2$ fields
triggered by the  breaking of the $SU(2)_R$
symmetry. However we may still ask whether the underlying
couplings, which determine the constants of proportionality, are
related by the $SU(2)_R$ symmetry, revealing a more symmetric
solution. Further relations between mass matrix elements require
a model of the underlying physics generating the higher dimension
operators and lie outside the scope of this paper.

Putting the constants of proportionality back in to the mass
matrices and dropping small elements gives the form

\begin{equation}
M_u\approx \left(
\begin{array}{ccc}
0 & b \epsilon^3 & 0 \\
b \epsilon^3 & \epsilon^2 &  a \epsilon \\
0 & a \epsilon & 1
\end{array}
\right)
\label{eq:muz}
\end{equation}

\begin{equation}
M_d\approx \left(
\begin{array}{ccc}
0 &  b' \bar{\epsilon}^3 & 0 \\
b' \bar{\epsilon}^3 & \bar{\epsilon}^2 & a' \bar{\epsilon} \\ 0
& a' \bar{\epsilon} & 1
\end{array}
\right)
\label{eq:8za}
\end{equation}
where $a, \; a, \; b$ and $b'$ are all of $O(1)$ and $SU(2)_R$
symmetry would imply $a=a' \ ; b=b'$.

Diagonalising these mass matrices gives the results of
eq(\ref{eq:tz}) and

\begin{eqnarray}
\frac{m_s}{m_b} & = & \mid \bar{\epsilon}^2 (1-a'^2) \mid
\nonumber \\
\frac{m_d m_s}{m_b^2} & = & \mid b'^2 \bar{\epsilon}^6 \mid
\nonumber \\
\frac{m_c}{m_t} & = & \mid \epsilon^2 (1-a^2) \mid \nonumber \\
\frac{m_u m_c}{m_t^2} & = & \mid b^2 \epsilon^6 \mid \nonumber
\\
\mid V_{cb} \mid &=& (a' \frac{m_s}{m_b}+ a \frac{m_c}{m_t}
+2\sqrt{a a' \frac{m_s m_c}{m_b m_t}} \cos{\phi})^{\frac{1}{2}}
\label{eq:results}
\end{eqnarray}
where $\phi$ and $\phi'$ are phases related to the phases of
$\epsilon, \; \bar{\epsilon}, \; a, \; b, \; a'$ and $b'$. The
predictions of eq(\ref{eq:tz}) for $ \mid V_{ub}/V_{cb}\mid $ and
$\mid V_{us} \mid$ follow from the texture zeros. The relation
for
$\mid V_{cb} \mid$ can be satisfied by a choice of $a'$ and $a$
of $O(1)$. Unfortunately it does not prove that the solution has
the larger $SU(2)_R$ symmetry with $a'=a$ although it is clearly
consistent with it. The quark mass ratios are successfully
predicted up to the coefficients of $O(1)$ because the
predictions
\begin{eqnarray}
(\frac{m_s}{m_b})^3\approx \frac{m_d m_s}{m_b^2} \nonumber \\
(\frac{m_c}{m_t})^3\approx \frac{m_u m_c}{m_t^2}
\label{eq:pred1}
\end{eqnarray}
hold for the observed quark masses (cf Table \ref{table:4} ). If
$SU(2)_R$ is a good symmetry we may derive from
eq(\ref{eq:results}) an equality (at the unification scale)
\begin{equation}
\frac{m_d m_b}{m_s^2}=\frac{m_u m_t}{m_c^2}
\label{eq:top}
\end{equation}
After including radiative corrections this is consistent with
present bounds.

Thus we conclude that the quark mass matrices are presently in
excellent agreement with a very large underlying symmetry. This
symmetry consists of left- right- symmetry together with a
$SU(2)_R$ symmetry of Yukawa couplings and a $U(1)$ horizontal
family symmetry. More detailed tests of this symmetry will be
forthcoming with the improved precision on the determination of
the quark mixing angles and the discovery of the top quark. As
is evident from Table \ref{table:4} the differences between the
various solutions are significant and should be measurable.

\section{Lepton masses}
Let  us now consider the structure of lepton masses resulting
from the $U(1)_{FD}$ symmetry. The lepton-antilepton charge
structure is given in eq(\ref{eq:pl}). If we are to maintain the
successful relation $m_b\approx m_{\tau}$ at the unification
scale we must have $\alpha_1=a_1$. In this case the lepton mass
matrix has the form
\begin{eqnarray}
M^L\approx \left(
\begin{array}{ccc}
\bar{\epsilon}^{\mid -4-2b\mid } &
\bar{\epsilon}^{\mid -3\mid } &
\bar{\epsilon}^{\mid -b-2\mid } \\
\bar{\epsilon}^{\mid -3\mid } &
\bar{\epsilon}^{\mid 2(b-1)\mid } &
\bar{\epsilon}^{\mid b-1\mid } \\
\bar{\epsilon}^{\mid -b-2\mid } &
\bar{\epsilon}^{\mid b-1\mid } & 1
\end{array}
\right)
\label{eq:mu0l}
\end{eqnarray}
where $b=a_2/a_1$. Although we have no measured lepton mixing
angles to guide us we will argue that even in the lepton sector
there {\it is} evidence for the {\it same} texture zero structure
as we have in the up and down quark matrices. The reason may be
seen from the structure of eq(\ref{eq:mu0l}) in which the (1,2)
and (2,1) matrix elements are independent of the parameter b and
are the same as the equivalent down quark matrix elements. This
means that if there are texture zeros in the (1,1) and (1,3)
positions (corresponding, as discussed above, to the range $b>1$)
we have the prediction $Det(M_d)=Det(M_l)$.  As originally
observed by Georgi and Jarlskog\cite{11}, this relation is in
excellent agreement with the observed masses when continued to
the unification scale for, together with the relation
$m_b=m_{\tau}$, it leads to the result
\begin{eqnarray}
m_d m_s & \approx & m_e m_{\mu} \; at \; the \; unification \;
scale
\nonumber \\
m_d m_s & \approx & 9m_e m_{\mu} \; at \; laboratory \; energies
\label{eq:det}
\end{eqnarray}
Thus the texture zero structure for leptons, which is predicted
for $b>1$ by the anomaly free structure of $U(1)_{FD}$, generates
the excellent mass prediction of eq(\ref{eq:det}).

To go further and determine $m_{\mu}$ and $m_e$ separately we
need to specify the value of b. For b=3, the lepton charges are
the same as the down quark sector, and so the structure of the
down quark and lepton mass matrices
are
identical.
 However there is another
choice of the    charge ratio b which does explain the
structure of this sector.    If b=3/2 the lepton mass matrix has
the form
\begin{equation}
M_L = \left (
\begin{array}{ccc}
\bar{\epsilon}^{5} & \bar{\epsilon}^{3} &
 0  \\
\bar{\epsilon}^{3} & \bar{\epsilon} &
 0  \\
0   &  0 &
  1
\end{array}
\right)
\label{eq:7l}
\end{equation}
The zeros in this mass matrix result because with this choice for
b, there is a residual $Z_2$ discrete gauge symmetry after $U(1)$
breaking by which the  electron
and muon fields get transformed by a factor $(-1)$.
In this case we have the relations at laboratory energies
\begin{eqnarray}
m_{\mu}\approx \frac{m_s}{3\bar{\epsilon}} \nonumber \\
m_e\approx \frac{m_d \bar{\epsilon}}{3}
\label{eq:mue}
\end{eqnarray}
which are in good agreement with the values of Table
\ref{table:4} for the value $\bar{\epsilon}=0.23$ needed to fit
the down quark masses and mixing angles.

\section{Anomaly cancellation and the structure of $U(1)_{FI}$}
\label{sec:final}

We have seen that a simple assignment of $U(1)_{FD}$ charges
leads to
predictions for the structure of the quark and lepton masses in
remarkable agreement with experiment. However we have also noted
that this is only possible if we assign charges to the Higgs
fields which apparently introduce an $SU(2)^2 U(1)$ anomaly. Here
we discuss cancellation of anomalies for the case that a
$U(1)_{FI}$ piece is added. We will show in this section that,
while there are still anomalies, they are of the type which may
be cancelled by the GS mechanism of string theory provided
$sin^2(\theta_W)=3/8$ at the string scale.

By definition $U(1)_{FD}$ in eq(\ref{eq:4}) acting on the quarks
and leptons is traceless and hence has vanishing $SU(3)^2U(1),\;
SU(2)^2U(1)$ and $U(1)_Y^2U(1)$ anomalies\footnote{We have use
the freedom to define the Higgs $U(1)$ charges to be entirely in
$U(1)_{FI}$.}. Thus for an anomaly free solution we must choose
$U(1)_{FI}$ to be anomaly free. With the minimal particle content
of the
MSSM,       the only $U(1)$s which are  anomaly free
 and flavour      independent are the weak
hypercharge itself and a symmetry         $U(1)_H$
which gives opposite charge to the two doublets.
As we noted at the end of Section \ref{sec:down} this does not
allow for viable mass matrices. However in theories derived from
a string theory there is a significant new
possibility for a non-vanishing anomaly associated with a new
$U(1)$ gauge factor can be cancelled by  the Green-Schwarz (GS)
anomaly cancellation
mechanism\cite{6,17}.

In the 4-D version of the GS mechanism one  cancels the
anomalies of a single $U(1)$ by an appropriate shift of
the axion present in the dilaton multiplet of four-dimensional
strings. This happens because such an axion has a direct coupling
to $F\tilde F$.
   For
the GS mechanism to be possible, the coefficients $A_i,\
i=3,2,1$ of the mixed anomalies of the
$U(1)$ with $SU(3)$, $SU(2)$ and $U(1)_Y$ have to be in the ratio
$A_3:A_2:A_1=k_3:k_2:k_1$ \cite{18} .  Here $k_i$ are the Kac-
Moody levels                  of the corresponding gauge factors
and they determine the                        boundary condition
of the gauge couplings at the string scale by                 the
well-known equation
$g_3^2k_3=g_2^2k_2=g_1^2k_1^2$. The usual                (e.g.
GUT) canonical values for these normalization factors
           (corresponding to the successful result
$sin^2(\theta_W)=3/8$)                   yield
$k_3:k_2:k_1=1:1:5/3$ and hence the GS mechanism can only
           work in this case if the mixed anomalies of the $U(1)$
with the                  SM gauge factors are in the ratio
$A_3:A_2:A_1=1:1:5/3$\cite{18}.
One can easily convince
oneself that there are only two $U(1)$ symmetries with anomalies
consistent with this ratio of gauge coupling constants, namely
$U(1)_X$                 and $U(1)_{XX}$ given in Table
\ref{table:2} .

\begin{table}
\begin{center}
\begin{tabular}{|c|ccccccc|}
\hline
   & Q & u & d &
L & e & $H_2$ & $ H_1$   \\
\hline
  $U(1)_H$ & 0 & 0 & 0 & 0 & 0 & 1 & -1
\\
\hline
  $U(1)_{XX}$ & 0 & 0 & 1 & 1 & 0 &  0&  0
\\
\hline
  $U(1)_{X}$ & 1 & 1 & 0 & 0 & 1 & 0 & 0
\\
\hline
\end{tabular}
\end{center}
\caption{Anomaly-free $U(1)_{FI}$
symmetries.}
\label{table:2}
\end{table}

\begin{table}
\begin{center}
\begin{tabular}{|c|ccccccc|}
\hline
   & Q & u & d &
L & e & $H_2$ & $ H_1$   \\
\hline
  $U(1)$ & $\alpha _i+x$ & $\alpha _i+x$ & $\alpha _i+y$
&                      $a_i+y$ & $a_i+x$ &  z-2$\alpha _1$ & -
z+$2\alpha _1$
\\
\hline
\end{tabular}
\end{center}
\caption{Anomaly-free $U(1)$
symmetries.}
\label{table:5}
\end{table}

 Thus, in a supersymmetric SM coming from a string the
most                       general family-independent
anomaly-free $U(1)$ consistent with                   canonical
gauge coupling unification is given
by:

\begin{equation}
U(1)_{FI}\  =\ z \ U(1)_H \ + \ x\ U(1)_X \ +\  y\ U(1)_{XX}\ .
\label{eq:4fi}
\end{equation}
          The full charges of
the $U(1)$ factor of eq(\ref{eq:4}) may now be  determined using
Tables \ref{table:2a} and \ref{table:2} and give    the charges
shown in Table
\ref{table:5}\footnote{The terms proportional to $2\alpha_1$ in
$H_{1,2}$ could be absorbed in z.}.

The choice $z=-2x$ gives the results of eq(\ref{eq:mu}) for the
up quark mass matrix. If one further has $3x+y=-4\alpha _1$  one
gets
the   results of eq(\ref{eq:8}) for the d-quark masses. Note,
however, the choice of the flavour-independent component allows
for further possibilities for the down quark matrices. In
particular, the alternative given by eq(\ref{eq:9})  may be
obtained if $3x+y=-7/2\alpha _1$. We conclude that the generic
problem raised by anomaly cancellation may naturally be solved
in the context of string based models.

So far we have assumed that $sin^2(\theta_W)=3/8$ at the string
unification scale. In fact an acceptable pattern of fermion
masses actually {\it requires} this value!  To see this let us
compute the mixed anomalies $A_i$ for the $U(1)_{FD}$ symmetry
of Table \ref{table:2a}  with $SU(3), SU(2)$ and $U(1)_Y$. One
finds respectively:

\begin{eqnarray}
A_3\ & =  & \  2\sum _{i=1}^3 \alpha _i\   \nonumber
\\
A_2\ & =  & \ {3\over 2}\sum _{i=1}^3 \alpha _i\ +\
{1\over 2} \sum _{i=1}^3 a_i\ +\ {\alpha _1\over 2}(w-2)
\nonumber \\
 A_1\ & = & \ {{11}\over 6}\sum _{i=1}^3 \alpha _i\ +\
{3\over 2} \sum _{i=1}^3 a_i\ +\
\alpha_1(w-2)
\label{eq:greens}
\end{eqnarray}
where, as we are working with the full $U(1)$ charges, we no
longer have $\alpha_3=-(\alpha_1+\alpha_2)$ and $a_3=-(a_1+a_2)$
as the $U(1)_{FI}$ piece adds a family independent term to
$\alpha_i$ and also to $a_i$. However, to maintain the result
$m_b=m_{\tau}$ we have $a_1=\alpha _1$. Furthermore, the texture
zeros together with the result $Det(M_L)=Det(M_d)$ requires
$a_2+a_3=\alpha _2+\alpha _3$. Thus the requirement of an
acceptable mass pattern gives the constraint:

\begin{equation}
\sum _{i=1}^3 \alpha _i\   =  \  \sum _{i=1}^3 a_i\
\label{eq:sum}
\end{equation}

{}From eq(\ref{eq:greens}) and eq(\ref{eq:sum}) we see that the
condition $A_3=A_2$, which is needed if the $SU(3)$ and $SU(2)$
couplings are unified, gives
\begin{eqnarray}
\\
w\ =\ 2\ \ \ or\ \alpha _1=0 \nonumber \\
and \; \; A_3:A_2:A_1=1:1:\frac{5}{3}
\label{eq:greecon}
\end{eqnarray}

The second equation requiring $sin^2(\theta_W)=3/8$ may be seen
as a {\it consequence} of an acceptable mass matrix structure.
The first condition in eq(\ref{eq:greecon}) requires $w=2$ since
$\alpha_1=0$ does not lead to an acceptable mass pattern.

 To summarise this section we have seen that a simple $U(1)_{FD}$
extension of the standard   model generates much of the quark and
lepton mass matrix   structure but that anomaly cancellation
requires a $U(1)_{FI}$ component. Moreover the full $U(1)$
extension, including the Higgs doublets      needed for the MSSM,
is only anomaly free in the context of string   theory via a
Green Schwarz term. The charges of each individual
particle with respect to this anomaly-free $U(1)$ giving rise to
the favoured solution eqs(\ref{eq:mu}), (\ref{eq:8}) and
(\ref{eq:7l}) are shown in Table 6.

\begin{table}
\begin{center}
\begin{tabular}{| c | ccccccccc |}\hline
$U(1)_F$ & Q & u & d & L & e & $H_1$ & $H_2$ & $\theta $ &
$\bar{\theta}$ \\
\hline
3rd generation & 0 & 0 & 0 & 0 & 0 & 0 & 0 & 1 & -1  \\
2nd generation & 1 & 1 & 1 & 1/2 & 1/2 & & & &  \\
1st generation &-4 &-4 &-4 &-7/2 &-7/2 & & & &  \\
\hline
\end{tabular}
\caption{Anomaly-free $U(1)$ gauge symmetry giving rise to the
textures in
eqs(7,11,21)}
\end{center}
\label{table:6}
\end{table}

The simplicity of the assignments is remarkable.
It is also worth emphasizing that this $U(1)$
symmetry may be made anomaly free through the GS mechanism {\it
if and only if the normalization of the coupling constants is the
canonical one $g_3^2=g_2^2=5/3g_1^2$} yielding the succesfull
prediction $sin^2\theta _W=3/8$. Thus the present scheme not only
predicts a succesfull pattern of fermion masses and mixings but
also predicts $sin^2\theta _W=3/8$ even without any grand
unification group.

   It is important
also to  recall  that
the $U(1)$s whose anomalies are cancelled through a
  GS mechanism are {\it necessarily spontaneously broken not far
below the string scale}. The reason for this is that the piece
 in the Lagrangian cancelling the anomalies has a
supersymmetric    counterpart which is a sort of field-dependent
Fayet-Iliopoulos  term for the $U(1)$. This term forces $U(1)$
symmetry breaking   in a natural way at a scale of order
$1/\sqrt{192} \pi M_{string}$   \cite{19,20}
Thus the present scheme also   explains why the extra $U(1)$
symmetry required to generate    the  fermion mass
patterns      does not survive
down to low energies.

We close this section with two comments about the consistency of
our solution with the other symmetries needed to build a viable
supersymmetric theory. The first concerns the $\mu$ problem. This
refers to the necessity to explain why the
Higgs scalars $H_1$ and $H_2$ needed in the MSSM are light even
though the mass term $H_1H_2$ is $SU(3)\otimes SU(2) \otimes
U(1)$ invariant and hence would naturally be expected to be
large. The problem can be solved by a discrete (gauge) symmetry
unbroken down to the electroweak scale. It is straightforward to
show that the smallest discrete symmetry capable of eliminating
these terms is a $Z_3$ symmetry which, however, necessarily is
not flavour blind in the lepton sector\cite{ir}. Indeed it allows
all possible Higgs couplings to
quarks but the (1,3) and (2,3) couplings to leptons are
forbidden. Clearly this symmetry is consistent with the form of
eq(\ref{eq:7l}) which followed from the choice $b=3/2$.

The second comment concerns the possible baryon and lepton number
violating terms. The $U(1)$ symmetry determining the
structure of masses does not forbid the presence of R-parity
violating terms like $t_Rb_Rss_R$ or $t_Lb_L^c\tau _L$, it was
not designed to do that. If such terms are to be suppressed there
must be an extra  discrete symmetry doing the job.
Alternatively, one can try to extend the above $U(1)$ to forbid
all R-parity violating terms. Adding to the $U(1)_{FI}$ symmetry
one piece proportional to the third component of right-handed
weak isospin  $U(1)_R$
may be useful in this respect.
If this is done, one  has to be careful so that a
residual discrete $Z_N$ gauge symmetry survives
doing the job after $U(1)$ symmetry breaking.

\section{Summary and Conclusions.}

To summarise, eqs(\ref{eq:mu}), (\ref{eq:8}) and
(\ref{eq:7l}) determine the order of magnitude of the 12 quark
and lepton masses and mixing angles in terms of just
four parameters, $\mid \epsilon \mid$, $\mid \bar{\epsilon} \mid$
together with the top Yukawa coupling, $h_t$ and the ratio of
Higgs vevs given in terms of the usual parameter $\tan \beta$.
Adding the requirement that the radiative correction should give
an acceptable b quark mass reduces this essentially to just three
for $m_t\ge 170Gev$ is required to keep $m_b$ from becoming too
large\cite{9e}. In addition, consistency with this fermion mass
structure yields the successful prediction $sin^2(\theta_W)=3/8$
at the unification scale.

If we wish to determing the precise magnitude of the 13 masses,
mixing angles and CP violating phase, the most symmetric solution
has the eight parameters of eqs(\ref{eq:muz}), (\ref{eq:8za}) and
(\ref{eq:7l}), $\mid 1-a^2 \mid$, $ \mid b \mid$, $\mid \epsilon
\mid$, $\mid \bar{\epsilon} \mid$, $\phi$, $\phi'$, together with
the top Yukawa coupling, $h_t$ and the ratio of Higgs vevs given
in terms of the usual parameter $\tan \beta$. This is consistent
with a very large underlying symmetry given by a horizontal U(1)
family symmetry constrained by anomaly cancellation, left- right-
symmetry giving symmetric mass
matrices, $SU(2)_R$ symmetry relating up and down quark couplings
and a down- quark lepton symmetry.

We find it quite remarkable that the
apparently complicated pattern of quark and lepton masses may be
explained by a very simple flavour and family symmetry of the
type discussed here.  The fact that a suitable choice of symmetry
plus multiplet structure is sufficient to fix a
phenomenologically realistic pattern of masses and mixing angles
demonstrates how the problem of understanding the latter may be
transformed into the problem of determining the former.  In 4-D
string theories the symmetries and multiplet structure are just
the things that are expected in a definite string
compactification.  What our analysis shows is that the symmetries
of the Yukawa couplings may be significant and point to an
underlying unification certainly consistent with the expectation
in string theory.  Of course the next step is to identify the
``correct'' 4-D string theory, but that is another story!


\begin{thebibliography}{99}


\bibitem{9e}
G. Costa, J. Ellis, G.L. Fogli, D.V. Nanopolous and F. Zwirner,
Nucl.Phys. B297   (1988) 244;J. Ellis, S. Kelley and D.V.
Nanopoulos, Phys. Lett. B249 (1990)441;Phys. Lett. B260 (1991)
131;
P. Langacker, Pennsylvania preprint UPR-0435T, (1990);
U. Amaldi, W. de Boer and H. F\"urstenau, Phys. Lett. B260 (1991)
447;P. Langacker and M. Luo, Phys.Rev.{\bf D44} (1991) 817;
G.G.~Ross and R.G. Roberts, \NPB{377}{92}{571}; F.Anselmo, L.
Ciafarelli, A. Peterman and A. Zichichi, Nuovo Cim. 104A (1991)
1817 and CERN preprint CERN-TH.6429/92.



\bibitem{9f} K. Inoue et al., Prog.Theor.Phys. {\bf 68} (1982)
927;
L.E. Ib\'a\~nez, Nucl.Phys. B218 (1983) 514;
L.E. Ib\'a\~nez and C. L\'opez, Phys. Lett. B126 (1983) 54;
Nucl.Phys. B233 (1984) 511;
L. Alvarez-Gaume, J. Polchinsky and M. Wise, Nucl.Phys. B221
(1983) 495; L.E Ib\'a\~nez, C. L\'opez and C. Mu\~noz, Nucl.
Phys. B256                  (1985)
218.

\bibitem{1} M.S.~Chanowitz, J.~Ellis and M.K.~Gaillard,
\NPB{128}{77}{506}.
 A.~Buras, J.~Ellis, M.K.~Gaillard and D.V.~Nanopoulos,
\NPB{135}{78}{66}.
%
\bibitem{2} H.~Arason, D.J.~Casta\~no, B.~Keszthelyi,
S.~Mikaelian, E.J.~Piard, P.~Ramond and B.D.~Wright,
\PRL{67}{91}{2933}. %

\bibitem{3} S.~Weinberg, in `` A Festschrift for I.I.~Rabi"
[Trans. N.Y. Acad. Sci., Ser. II (1977), v. 38], p. 185;
F.~Wilczek and A.~Zee, \PLB{70}{77}{418}.
T. Maehara and T. Yanagida, Prog. Theor. Phys. 60 (1978) 822 J.
Chakrabarti, Phys. Rev. D20 91979) 2411
F. Wilczek and A. Zee, Phys. Rev. Lett. 42 (1979) 421
%
\bibitem{4} H.~Fritsch,\PLB{70}{77}{436}; \PLB{73}{78}{317};
F.J.~Gilman and Y.~Nir, \ARNP{40}{90}{213};
 P.~Kaus and S.~Meshkov, Mod. Phys. Lett. {\bf A3}(1988) 1251.
%

\bibitem{5}

C.D.Froggat and H.B. Nielsen, Origin of symmetries, World
Scientific (1991)
N. Cabibbo, Phys. Rev. Lett. 10 (1963) 531
M. Kobayashi and T. Maskawa, Prog. Theor. Phys. 49 (1973) 652
M.E. Machacek and M.T. Vaughn, Phys. Lett. B103 (1981) 427
\bibitem{5e} J.~Harvey, P.~Ramond and D.~Reiss,
\PLB{92}{80}{309}; S.~Dimopoulos, L.J.~Hall and S.~Raby,
\PRL{68}{92}{1984}; \PRD{45}{92}{4195}; H.~Arason,
D.J.~Casta\~no, P.~Ramond and E.J.~Piard,
\PRD{47}{93}{232}; G.F.~Giudice, Mod. Phys. Lett. {\bf A7} (1992)
2429. %


\bibitem{6}For a review of string theories, see M. Green, J.
Schwarz and E. Witten, Superstring Theory, Cambridge University
Press, 1987.

\bibitem{e1} A.~Buras and M.K.~Harlander, Munich preprint MPI-
PAE/PTh 1/92;
J.L.~Rosner, Jour. Phys. {\bf G18}(1992) 1575.


\bibitem{7} P.Ramond, R.G.Roberts and G.G.Ross, Nucl. Phys.
B406(1993)19

\bibitem{8}C.D.Froggart and H.B.Nielsen, ``Origin of
Symmetries'', World Scientific 1991.

\bibitem{9}M.Leurer, Y.Nir and N.Seiberg,
Nucl.Phys.B398(1993)319;Rutgers preprint RU-93-43, WIS-93/93/Oct-
PH(1993)
Y.Nir and N.Seiberg, Phys.Lett. B309(1993)337

\bibitem{10}

C.D. Froggatt and H.B. Nielsen, Nucl. Phys. B147 91979) 277 C.D.
Froggatt and H.B. Nielsen, Nucl. Phys. B164 (1979) 144 S.
Dimopoulos, Phys. Lett. B129 (1983) 417
G. Anderson et al, lawrence Berkeley Lab Preprint LBL-33531, UCB-
PTH-93/03 (1993)
\bibitem{10e} H.~Fritzsch and J.~Plankl,\PLB{237}{90}{451} and
references  therein.

\bibitem{11} H.~Georgi and C.~Jarlskog, \PLB{86}{79}{297}.  %

 \bibitem{17} M. Green and J. Schwarz, \PLB{149}{84}{117}

\bibitem{18} L.E. Ib\'a\~nez, \PLB{303}{93}{55}

\bibitem{19} M. Dine, N. Seiberg and E. Witten, Nucl.Phys. B289
(1987) 585; J.Atick, L. Dixon and A. Sen, Nucl.Phys. B292 (1987)
109; M. Dine, I. Ichinoise and N. Seiberg, Nucl.Phys. B293 (1987)
253.

\bibitem{20} A. Font, L.E. Ib\'a\~nez, H.P. Nilles and F.
Quevedo, Nucl.Phys. B307 (1988) 109; \PLB{210}{88}{101};
J.A. Casas, E.K. Katehou and C. Mu\~noz, Nucl.Phys. B317 (1989)
171; J.A. Casas and C. Mu\~noz, \PLB{209}{88}{214}
;\PLB{214}{88}{63}; A. Font, L.E. Ib\'a\~nez, F. Quevedo and A.
Sierra, Nucl.Phys. B331 (1990) 421.



\bibitem{ir}L.E.Ibanez and G.G.Ross, in preparation.
\end{thebibliography}
\end{document}